\documentclass[%
 aip,
rsi,%
 amsmath,amssymb,
reprint,%
]{revtex4-1}

\usepackage{graphicx}
\usepackage{dcolumn}
\usepackage{bm}
\usepackage{fixltx2e}
\usepackage[utf8]{inputenc}
\usepackage[T1]{fontenc}
\usepackage{mathptmx}
\usepackage{graphicx}
\usepackage{float}
\usepackage{color}
\usepackage[normalem]{ulem} 

\begin{document}
\preprint{AIP/123-QED}
\title{Anisotropic dielectric functions, band-to-band transitions, and critical points in $\alpha$-Ga$_2$O$_3$ }
\author{Matthew Hilfiker}
\email{mhilfiker2@unl.edu}
\homepage{http://ellipsometry.unl.edu}
\affiliation{Department of Electrical and Computer Engineering, University of Nebraska-Lincoln, Lincoln, NE 68588, USA}
\author{Rafa\l{} Korlacki}
\affiliation{Department of Electrical and Computer Engineering, University of Nebraska-Lincoln, Lincoln, NE 68588, USA}
\author{Riena Jinno}
\affiliation{School of Electrical and Computer Engineering, Cornell University, Ithaca, NY 14853, USA}
\affiliation{Department of Electronic Science and Engineering, Kyoto University, Kyoto, 615-8510, Japan}
\author{Yongjin Cho}
\affiliation{School of Electrical and Computer Engineering, Cornell University, Ithaca, NY 14853, USA}
\author{Huili Grace Xing}
\affiliation{School of Electrical and Computer Engineering, Cornell University, Ithaca, NY 14853, USA}
\affiliation{Department of Material Science and Engineering, Cornell University, Ithaca, NY 14853, USA}
\author{Debdeep Jena}
\affiliation{School of Electrical and Computer Engineering, Cornell University, Ithaca, NY 14853, USA}
\affiliation{Department of Material Science and Engineering, Cornell University, Ithaca, NY 14853, USA}
\author{Ufuk Kilic}
\affiliation{Department of Electrical and Computer Engineering, University of Nebraska-Lincoln, Lincoln, NE 68588, USA}
\author{Megan Stokey}
\affiliation{Department of Electrical and Computer Engineering, University of Nebraska-Lincoln, Lincoln, NE 68588, USA}
\author{Mathias Schubert}
\affiliation{Department of Electrical and Computer Engineering, University of Nebraska-Lincoln, Lincoln, NE 68588, USA}
\affiliation{Terahertz Materials Analysis Center and Center for III-N technology, C3NiT -- Janz\`{e}n, Department of Physics, Chemistry and Biology (IFM), Link\"{o}ping University, 58183 Link\"{o}ping, Sweden}
\affiliation{Leibniz Institut f\"{u}r Polymerforschung e.V., 01069 Dresden, Germany}

\date{\today}

\begin{abstract}
We use a combined generalized spectroscopic ellipsometry and density functional theory approach to determine and analyze the anisotropic dielectric functions of an $\alpha$-Ga$_2$O$_3$ thin film. The sample is grown epitaxially by plasma-assisted molecular beam epitaxy on $m$-plane sapphire. Generalized spectroscopic ellipsometry data from multiple sample azimuths in the spectral range from 0.73~eV to 8.75~eV are simultaneously analyzed. Density functional theory is used to calculate the valence and conduction band structure. We identify, for the indirect-bandgap material, two direct band-to-band transitions with $M_0$-type van Hove singularities for polarization perpendicular to the $c$ axis, $E_{0,\perp}=5.46(6)$~eV and $E_{0,\perp}=6.04(1)$~eV, and one direct band-to-band transition with $M_1$-type van Hove singularity for polarization parallel with $E_{0,||}=5.44(2)$~eV. We further identify excitonic contributions with small binding energy of 7~meV associated with the lowest ordinary transition, and a hyperbolic exciton at the $M_1$-type critical point with large binding energy of 178~meV.
\end{abstract}

\maketitle

Ga$_2$O$_3$ has received recent research interest as an ultra-wide bandgap semiconductor due to a predicted breakdown electric field more than double of SiC and GaN.\cite{HigashiwakiAPL2018Guest} Of the five polymorphs of Ga$_2$O$_3$ ($\alpha$, $\beta$, $\gamma$, $\delta$, and $\epsilon$)\cite{RoyJACS1952}, the metastable form of $\alpha$-Ga$_2$O$_3$ has been of recent interest because it can be formed and stabilized using heteroepitaxial growth on sapphire ($\alpha$-Al$_2$O$_3$) substrates at low-temperatures.\cite{ShinoharaJJAP} $\alpha$-Ga$_2$O$_3$ is thus of interest because it can be grown on widely available substrates, such as synthetic sapphire. Furthermore, it has a more favorable symmetry (rhombohedral, $R \bar{3}c$, space group 167) than the thermodynamically stable version $\beta$-Ga$_2$O$_3$ (monoclinic) which may result in simpler device design and more reliable performance characteristics. Knowledge of band structure properties and band-to-band transitions is of fundamental importance for potential application in semiconductor devices. Optical investigations permit access to direct fundamental bandgap and higher-energy transitions, including their polarization characteristics, in optically uniaxial semiconductors for electric field directions parallel ($\parallel$, extraordinary direction) and perpendicular ($\bot$, ordinary direction) to the lattice $c$ axis. Multiple studies have been performed to identify the bandgap energies and parameters for optical transitions in $\alpha$-Ga$_2$O$_3$ and consistent answers have not yet emerged. Sinha~\textit{et al.} studied $\alpha$-Ga$_2$O$_3$ nanocrystalline thin films obtained with the sol-gel technique and estimated an isotropically polarization averaged bandgap of 4.98~eV using transmission measurements.\cite{SinhaaGO} Shinohara~\textit{et al.} estimated a bandgap energy of 5.3~eV from transmission measurements of heteroepitaxially grown $\alpha$-Ga$_2$O$_3$ thin films using ultrasonic mist chemical vapor deposition.\cite{ShinoharaJJAP} Roberts~\textit{et al.} reported an optical bandgap between 5.15 $\dots$ 5.2~eV from transmission measurements for $\alpha$-Ga$_2$O$_3$ thin films grown using low temperature plasma-enhanced atomic layer deposition.\cite{RobertsJCGaGO} Segura~\textit{et al.} reported near-bandgap spectral range transmission intensity and reflection-type spectroscopic ellipsometry measurements on $c$-plane $\alpha$-Ga$_2$O$_3$ thin films deposited by ultrasonic mist chemical vapor phase epitaxy.\cite{SeguraPRMataGOEg2017} Two prominent absorption peaks were identified for the ordinary direction and ascribed to allowed direct transitions from crystal-field split valence bands to the same conduction band. Excitonic effects with large Gaussian broadening were modelled using the Elliott-Toyozawa approach,\cite{ElliottPRB1957,ToyozawaPTPExciton1958} but due to large broadening the available data was too limited in spectral range ($\approx$~6.5~eV) to make decisive conclusions about the nature of the band-to-band transitions in $\alpha$-Ga$_2$O$_3$. A large exciton binding energy of 110~meV was suggested and band-to-band transitions of $E_{0,\bot}$=5.61~eV and $E_{1,\bot}$=6.44~eV were reported. Kracht~\textit{et al.} investigated $r$-plane oriented $\alpha$-Ga$_2$O$_3$ thin films grown by plasma-assisted molecular beam epitaxy (PAMBE).\cite{KrachtPRApplaGO2018} In this crystallographic orientation, the lattice $\mathbf{c}$ direction is not parallel to the thin film normal and sensitivity to $\varepsilon_{\bot}$ and $\varepsilon_{\parallel}$ can be obtained by aligning the sample once with the $c$ axis and once with the $[1\bar{1}02]$ direction parallel to the plane of incidence, respectively. An ad-hoc lineshape analysis approach suggested band-to-band transition energies of $E_{0,\bot}$=5.62~eV and $E_{0,\parallel}$=5.58~eV. $E_{1,\bot}$ could only be guessed within 6.18 $\dots$ 6.52~eV. Exciton binding energies were found to be 38~meV for both directions. The high frequency dielectric constants were extrapolated to $\varepsilon_{\infty,\perp}$=3.75 and $\varepsilon_{\infty,\parallel}$=3.64.  Feneberg~\textit{et al.} performed an ultra wide range spectroscopic ellipsometry study from the infrared (40~meV) to the vacuum ultra violet (20~eV) on $c$-plane oriented $\alpha$-Ga$_2$O$_3$ thin films.\cite{FenebergPRMataGOoDF2018} This work found $E_{0,\bot}$=5.8~eV, and $E_{1,\bot}$ was not observed. A higher energy transition was observed at $E_{2,\bot}$=11.1~eV. Feneberg~\textit{et al.} measured $\varepsilon_{\bot}$ and $\varepsilon_{\parallel}$ in the infrared spectral regions on $m$-plane $\alpha$-Ga$_2$O$_3$ thin films grown by mist chemical vapor epitaxy. The authors performed phonon mode analysis and reported 5 out of 6 infrared active modes. Using the approach described in Ref.~\onlinecite{KasicPRB62_2000} the authors determined the conduction band effective mass parameters of $m^{*}_{\bot}$ = (0.297$\pm$0.010)$m_{\mathrm{e}}$ and $m^{*}_{\parallel}$ = (0.316$\pm$0.007)$m_{\mathrm{e}}$.\cite{FenebergAPLaGOIR2019} A first-principles density functional theory (DFT) all-electron basis calculation was reported by He~\textit{et al.} presenting dielectric function, reflectance, and energy-loss function up to 50~eV, and an electron effective mass parameter ($m^*_{\mathrm{e}}$=0.276$m_{\mathrm{e}}$).\cite{PhysRevB.74.195123} Litimein~\textit{et al.} employed full-potential linearized augmented plane-wave method and reported density of states and anisotropic dielectric function of $\alpha$-Ga$_2$O$_3$.\cite{LITIMEIN2009148} Furthm\"uller and Bechstedt used a DFT approach and Bethe-Salpeter equation calculations and presented band structure, density of states and anisotropic dielectric functions, and reported a bandgap of 5.63~eV.\cite{PhysRevB.93.115204} Bechstedt and Furthm\"{u}ller calculated the effective mass anisotropy and predicted an isotropically averaged exciton binding energy of 184~meV.\cite{doi:10.1063/1.5084324}

Direct band-to-band transitions cause critical point (CP) structures in the dielectric functions, $\varepsilon_{\bot,\parallel}$, which possess unique frequency characteristics depending on the type of the associated singularity in the combined density of states (van Hove singularity).\cite{Yu99} Crucial for accurate lineshape analysis of $\varepsilon_{\bot,\parallel}$ is detailed knowledge of band structure properties. Furthermore, band-to-band transitions are accompanied by excitonic contributions, which also depend on the behavior of the participating bands.\cite{Yu99} A lineshape analysis using CP structures and comparison with band structure calculations has not previously been performed for $\alpha$-Ga$_2$O$_3$ and is reported here using results of a combined generalized spectroscopic ellipsometry (GSE) and DFT analysis approach. In our sample the $c$ axis is parallel to the surface and permits accurate measurements of $\varepsilon_{\parallel}$ and $\varepsilon_{\bot}$ from simultaneous analysis of data measured at multiple sample azimuths in the spectral range from 0.73~eV to 8.75~eV. We perform DFT calculations and obtain the valence and conduction band structure in $\alpha$-Ga$_2$O$_3$, which indicate an indirect-bandgap material. We further identify the origins of the singularities in the combined density of states, which lead to $M_0$-type CP structures at the band-to-band transitions in $\varepsilon_{\bot}$, and to one $M_1$-type CP structure in $\varepsilon_{\parallel}$. We identify excitons with small binding energy associated with the $M_0$-type transitions, and hyperbolic excitons\cite{Masaki2DJPSJ1966,Yu99} with large exciton binding energy for the $M_1$-type transition. We compare and discuss our findings with previous observations. We provide complete parameter sets for $\varepsilon_{\bot,\parallel}$ which will become useful for ellipsometric model analysis of heterostructures containing $\alpha$-Ga$_2$O$_3$ thin films.

DFT calculations were performed using the plane-wave code Quantum ESPRESSO\cite{[{Quantum ESPRESSO is available from http://www.qu\-an\-tum-es\-pres\-so.org. See also: }]GiannozziJPCM2009QE} with a combination of generalized-gradient-approximation (GGA) density functional of Perdew, Burke and Ernzerhof\cite{PhysRevLett.77.3865} and norm-conserving Troullier-Martins pseudopotentials originally generated using FHI98PP\cite{FuchsCPC1999,PhysRevB.43.1993} available in the Quantum ESPRESSO pseudopotentials library. The pseudopotential for gallium did not include the semicore $3d$ states in the valence configuration. All calculations were performed with a high electronic wavefunction cutoff of 400 Ry. As the starting point we used structural parameters from the Materials Project.\cite{Jain2013, osti_1188790} 
The calculations were performed in a rhombohedral cell:
\begin{align*}
    \mathbf{p}_1 &= (a_H \sqrt{3}/3, 0, c_H/3),\\
    \mathbf{p}_2 &= (-a_H \sqrt{3}/6, a_H/2, c_H/3),\\
    \mathbf{p}_3 &= (-a_H \sqrt{3}/6, -a_H/2, c_H/3),
\end{align*}
where $a_H$ and $c_H$ are parameters of the hexagonal cell. The initial structure was first relaxed to force levels less than 10$^{-6}$ Ry Bohr$^{-1}$. A dense shifted $8 \times 8 \times 8$ Monkhorst-Pack\cite{Monkhorst1976} grid was used for sampling of the Brillouin zone and a convergence threshold of $1 \times 10^{-12}$ Ry was used to reach self-consistency. The lattice parameters for the fully relaxed structure were $a_H=5.0101$ \AA~and $c_H=13.493$ \AA, similar to the values reported in the literature at the DFT/GGA level.\cite{PhysRevB.93.115204} In order to improve the quality of the DFT pseudo-wavefunction and bring the value of the bandgap closer to the experimentally measured one, we performed additional calculations using the hybrid Gau-PBE\cite{song2011,song2013} density functional. All the hybrid calculations were performed at the PBE equilibrium geometry using a regular non-shifted $8 \times 8 \times 8$ Monkhorst-Pack grid for the Brillouin zone sampling and $4 \times 4 \times 4$ grid for sampling of the Fock operator. The convergence threshold for self-consistency in hybrid functional calculations was $5 \times 10^{-9}$ Ry.

In order to study the band structure we used the band interpolation method based on the maximally localized Wannier functions\cite{PhysRevB.56.12847,PhysRevB.65.035109} as implemented in the software package WANNIER90\cite{mostofi2008}. The initial projectors for the Wannier functions were automatically generated using the selected columns of the density matrix (SCDM)\cite{vitale2020} method. For $\alpha$-Ga$_2$O$_3$ the lowest four conduction bands are not entangled with higher conduction bands allowing us to treat the valence band and the four lowest conduction bands together as an isolated system. As a result of the wannierisation procedure, with the convergence threshold set at $1.0 \times 10^{-12}$ \AA$^2$, we obtained a set of 28 maximally localized Wannier functions with an average spread of 1.03 \AA$^2$. These were then used to obtain a high resolution interpolated band structure. Finally, the allowed optical transitions at the Brillouin zone center were obtained by extracting matrix elements of the momentum operator between the valence and conduction bands, $|\mathcal{M}_{cv}|^2$. All non-trivial values of the matrix elements in the energy range below 10 eV are listed in Table~\ref{Tab:DFT}.

Heteroepitaxial $\alpha$-Ga$_{2}$O$_3$ films were grown on polished $m$-plane $\alpha$-Al$_{2}$O$_3$ substrates at thermocouple substrate temperature ($T$) of 650 $^\circ$C using PAMBE. Substrates received an oxygen plasma treatment prior to growth in the chamber at $T$ = 800 $^\circ$C for 10 min. During deposition, an oxygen flow rate of 0.5 sccm was introduced to create active oxygen species using a radio frequency plasma source (RF$_{power}$ = 250 W). Pressure was maintained during the growth process at ~10$^{-5}$ Torr. The epitaxial layer thickness was determined to be 51.8~nm from x-ray reflectivity measurements. Atomic force microscopy indicated a small root mean square roughness of 0.96~nm. The $\alpha$-Ga$_{2}$O$_3$ film was determined to be completely lattice-relaxed by asymmetrical reciprocal space map analysis. Growth and structural characterization are discussed in further detail by Jinno~\textit{et al.}\cite{jinno2020crystal}

Measurements of GSE data were performed at ambient temperature for the spectral range of 0.73 to 8.75 eV. A dual-rotating compensator ellipsometer (RC2, J. A. Woollam Co., Inc.) was used to acquire data in the spectral range of 0.73~eV to 6.42~eV at three angles of incidence ($\Phi_a$~=~50$^\circ$, 60$^\circ$, 70$^\circ$) for a full azimuthal rotation in steps of 15$^\circ$. A rotating-analyzer ellipsometer with an automated compensator function (VUV-VASE, J.A. Woollam Co., Inc.) to measure data in the vacuum-ultra-violet (VUV) spectral region. The VUV-VASE measured data for the 5~eV to 8.75~eV spectral region with a spectral resolution equal to 0.04~eV. Data was acquired at three angles of incidence ($\Phi_a$~=~50$^\circ$, 60$^\circ$, 70$^\circ$), and at azimuthal rotations in steps of 45$^\circ$.

We model the optical properties of $\alpha$-Ga$_2$O$_3$ using a uniaxial dielectric tensor with two major dielectric functions, $\varepsilon_{\bot}$, and $\varepsilon_{\parallel}$. To account for nanoscale thin film surface roughness, an effective medium approximation is applied with an ultra thin optical layer above $\alpha$-Ga$_2$O$_3$. An isotropic average of $\varepsilon_{\bot}$ and $\varepsilon_{\parallel}$ is weighted with 50$\%$ void ($\varepsilon_{void}$=1).\cite{Fujiwara_2007} Then a Cauchy dispersion equation is used to approximate $\varepsilon_{\bot}$ and $\varepsilon_{\parallel}$ in the below-bandgap region to determine the thickness of the roughness overlayer, the $\alpha$-Ga$_2$O$_3$ thickness, and the Euler angles for the sample which determine the orientation of $\mathbf{c}$, common to both thin film and substrate, during each experiment. These values are then fixed for the remainder of the analysis. The Cauchy equation is applied in the transparent region ($\hbar\omega \leq$ 4.5~eV) where negligible absorption does not affect the GSE data. Then, a point-by-point (PBP) regression analysis is performed for the full data set to determine simultaneously $\varepsilon_{\bot}$ and $\varepsilon_{\parallel}$. Initially, $m$-plane sapphire is analyzed to provide accurate optical constants of $\alpha$-Al$_{2}$O$_3$ for our subsequent analysis of $\alpha$-Ga$_{2}$O$_3$. Harman~\textit{et al.} determined the anisotropic optical constants of sapphire using spectroscopic ellipsometry in the spectral range of 0 to 30 eV. They found a fundamental absorption edge at $\sim$9 eV and determined $\varepsilon_{\infty,\bot}$ = 3.064 and $\varepsilon_{\infty,\parallel}$ = 3.038.\cite{doi:10.1063/1.357922} This is in excellent agreement with our experimental results of $\varepsilon_{\infty,\bot}$ = 3.068 and $\varepsilon_{\infty,\parallel}$ = 3.042.

A CP model dielectric function (MDF) approach is used to analyze $\varepsilon_{\bot}$ and $\varepsilon_{\parallel}$. As will be discussed below, we find that the two lowest transitions found in $\varepsilon_{\bot}$ each require a $M_0$-type CP structure\cite{Yu99,PhysRevB.60.16618,tompkins2006wvase32}

\begin{equation}
\varepsilon_{(M_0)} = AE^{-1.5}{\chi^{-2}[2 - (1 + \chi)^{0.5} - (1 - \chi)^{0.5}]},
\end{equation}

\begin{equation}
\chi = \frac{(\hbar\omega + i\Gamma)}{E},
\end{equation}

\noindent where $A$, $E$, and $\Gamma$, respectively, denote CP amplitude, transition energy, and broadening parameter, and $\hbar\omega$ is the photon energy. For the lowest transition in $\varepsilon_{\parallel}$, a $M_1$-type CP structure is needed\cite{StroessnerPRB1985,tompkins2006wvase32}

\begin{equation}
\varepsilon_{(M_1)} =- \frac{A}{\chi^2} ln[1 - \chi^2].
\end{equation}

\noindent This CP structure represents a van Hove singularity where the joint density of states reflects a saddle point with one of the combined effective mass parameters negative, or approaching zero.\cite{StroessnerPRB1985} For the two lowest transitions in both $\varepsilon_{\bot}$ and $\varepsilon_{\parallel}$, we observe strong excitonic contributions. We describe these with an anharmonically broadened Lorentz oscillator

\begin{equation}
    \varepsilon_{(ex)}=\frac{A^2-ib\hbar\omega}{E^2-(\hbar\omega)^2-i\Gamma\hbar\omega},\label{eq:anharmonicLor}
\end{equation}

\noindent where $b$ denotes the anharmonic broadening parameter.\cite{PhysRevB.99.184302}$^,$\footnote{A detailed discussion of the presentation of the anharmonically broadened oscillator form suggested by Gervais and Piriou in Ref.~\onlinecite{Gervais_1974} and the form suggested by Mock~\textit{et al.} in Ref.~\onlinecite{PhysRevB.99.184302} has yet to appear in the literature. Briefly, simple mathematical transformations proof that both forms are identical. The advantage of the latter form, Eq.~\ref{eq:anharmonicLor} in this paper, is that it can be added to a sum of model contributions to the dielectric function, while the former must be brought into a product form considerably complicating the regression calculations.} We note that excitonic contributions to the $M_0$-CP are interpreted as due to ground state contributions from three-dimensional effective hydrogen atom-like excitons, while excitonic contributions to the $M_1$-CP are interpreted as two-dimensional effective hydrogen atom-like (a.k.a. hyperbolic) excitons.\cite{Masaki2DJPSJ1966,Yu99} The Tanguy-Elliott model,\cite{ElliottPRB1957,TanguyPRL1995} which also includes exciton continuum states, did not suffice to provide a good match to our experimental data and was therefore not used. CP contributions from transitions at higher energies often contain contributions from multiple, neighboring energy transitions within the Brillouin zone which thus appear broadened and difficult to differentiate. We use a Gaussian broadened oscillator here for the imaginary ($\Im$) part

\begin{equation}
\Im\{\varepsilon_{(G)}\}= A\left(e^{-\left[\frac{\hbar\omega-E}{\sigma}\right]^2}-e^{-\left[\frac{\hbar\omega+E}{\sigma}\right]^2}\right), \label{gaussiane2}
\end{equation}
\begin{equation}
\sigma = \Gamma/(2\sqrt{ln(2)}),
\end{equation}

\noindent where the real part is obtained from Kramers-Kronig integration\cite{tompkins2006wvase32,Fujiwara_2007}

\begin{equation}
\Re\{\varepsilon_{(G)}\}= \frac{2}{\pi}P\int_0^\infty \frac{\xi\Im\{\varepsilon_{(G)}\}}{\xi^2-(\hbar\omega)^2}d\xi.\label{gaussiane1}
\end{equation}


\begin{figure}[htbp]
\centering
\includegraphics[width=\linewidth]{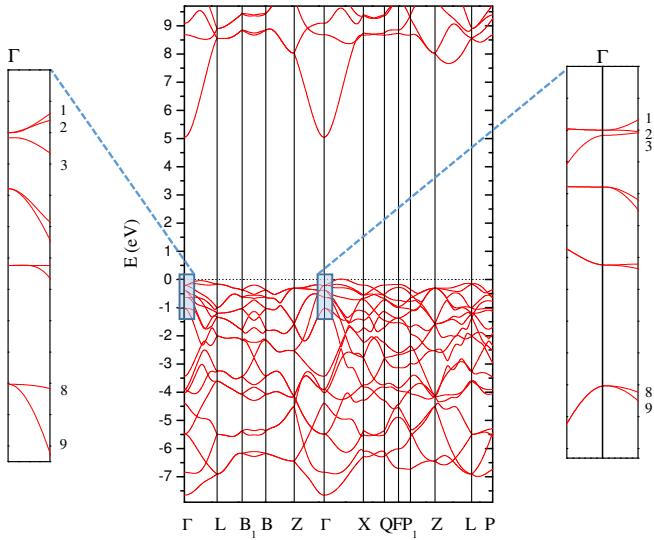}
\caption{Hybrid-level-DFT structure of the valence and conduction bands of $\alpha$-Ga$_2$O$_3$. The valence band maximum is set at $E=0$. The breakouts enlarge the top valence bands at the $\Gamma$-point, and bands involved in the band-to-band transitions observed in this work are indexed. See also Tab.~\ref{Tab:DFT}. Definition of high-symmetry points as in Ref.~\onlinecite{setyawan2010}. (Brillouin zone for the type-1 rhombohedral lattice, RHL$_1$, with the following variables: $\alpha = 55.868^{\circ}$, $\eta = 0.764395$, and $\nu=0.367803$. For definitions of $\eta$ and $\nu$, see Ref.~\onlinecite{setyawan2010}.)}
\label{fig:aGObandstructure}
\end{figure}

\begin{table}[ht]
\centering
\caption{Hybrid-level-DFT $\Gamma$-point direct band-to-band transition energies, $\Gamma_{c-v}$, and transition matrix elements, $|\mathcal{M}_{cv}|^2$ along ordinary (upper section) and extraordinary (lower section) directions. Indices are labeled from the bandgap, i.e., for conduction bands increasing with energy starting with $c = 1$ for the lowest conduction band, and for valence bands decreasing in energy starting with $v = 1$ for the highest band. }
\begin{tabular}{{l}{c}{c}{c}{c}{c}}
\hline \hline
Label &{$E$ (eV)} &{$|\mathcal{M}_{cv}|^2$ ($\hbar$/Bohr)$^2$}&{$c$}&{$v$}\\
\hline
$\Gamma_{1-1}$+$\Gamma_{1-2}$&5.250&0.24110&1&1,2\\
$\Gamma_{1-8}$+$\Gamma_{1-9}$&6.052&0.22577&1&8,9\\
$\Gamma_{1-11}$+$\Gamma_{1-12}$&8.953&0.00412&1&11,12\\
$\Gamma_{2-4}$+$\Gamma_{2-5}$&9.064&0.01284&2&4,5\\
$\Gamma_{2-6}$+$\Gamma_{2-7}$&9.309&0.06575&2&6,7\\
$\Gamma_{3-4}$+$\Gamma_{3-5}$&9.468&0.45691&3&4,5\\
$\Gamma_{3-6}$+$\Gamma_{3-7}$&9.713&0.00276&3&6,7\\
\hline 
$\Gamma_{1-3}$&5.266&0.20734&1&3\\
$\Gamma_{1-13}$&9.067&0.00718&1&13\\
\hline\hline
\label{Tab:DFT} 
\end{tabular}
\end{table}

\begin{table}[ht]
\centering
\caption{Hybrid-level-DFT $\Gamma$-point band effective mass parameters determined in directions $\Gamma - \mathrm{L}$, $\Gamma - \mathrm{Z}$, and $\Gamma - \mathrm{X}$ in units of free electron mass, $m_e$. Direction $\Gamma$-X is in-plane perpendicular to the hexagonal $c$ axis; $\Gamma$-Z is parallel to $c$, hence, their respective (conduction and valence band) effective mass parameters correspond to $m_{\perp}$ and $m_{||}$, respectively. Note that valence bands 1 and 2 are degenerate at the $\Gamma$-point. The reduced effective mass parameter is 0.18$m_{\mathrm{e}}$ for the ordinary transition ($c=1\leftarrow\rightarrow \nu=1,2$), and 0.14$m_{\mathrm{e}}$ for the extraordinary transition ($c=1\leftarrow\rightarrow\nu=3$).}
\begin{tabular}{{l}{c}{c}{c}{c}}
\hline \hline
{Band index} &{$m_{\Gamma - \mathrm{L}}/m_{\mathrm{e}}$} &{$m_{\Gamma - \mathrm{Z}}/m_{\mathrm{e}}$}&{$m_{\Gamma - \mathrm{X}}/m_{\mathrm{e}}$}\\
\hline
$c=1$& 0.28& 0.29& 0.28\\
\hline 
$v=1$& 0.70& 5.6& 0.51\\
$v=2$& 2.4& 5.6& >10\\
$v=3$& 3.2& -0.27& 1.4\\
$v=8$& -5.7& -0.18& -0.90\\
$v=9$& -0.23& -0.18& -0.38\\
\hline\hline
\label{tab:mass} 
\end{tabular}
\end{table}

Figure~\ref{fig:aGObandstructure} depicts the band structure of $\alpha$-Ga$_2$O$_3$ obtained from our DFT calculations using the hybrid density functional described above. $\alpha$-Ga$_2$O$_3$ is indirect, with valence band maximum outside of the zone center. Table~\ref{Tab:DFT} lists Gau-PBE hybrid functional obtained band-to-band transitions in the lowest energy region along with the transition matrix elements and bands involved. Two transitions nearly equal in amplitude are found for polarization perpendicular to the $c$ axis at 5.25~eV and 6.052~eV, and one for polarization parallel at 5.266~eV. The two lowest transitions involve different valence bands. A pronounced group of transitions is noted at approximately 9~eV for $\varepsilon_{\bot}$, and a rather weak transition in this range for $\varepsilon_{\parallel}$. In Table~\ref{tab:mass} we list the direction dependent effective mass parameters for all conduction and valence bands involved in the lowest three transitions. The effective mass parameters were obtained by fitting a second degree polynomial to the respective bands in the range of $\pm$~0.01~\AA$^{-1}$ from the $\Gamma$-point. We note that our conduction band parameters ($m_{\bot}$=0.28$m_{\mathrm{e}}$, $m_{\parallel}$=0.29$m_e$) are in good agreement with data reported by Feneberg~\textit{et al.}\cite{FenebergAPLaGOIR2019} When calculating the reduced mass parameters for the lowest transitions, one can observe that transitions perpendicular axis $c$ correspond to $M_0$-type singularities, while the transition parallel $c$ is of $M_1$-type due to the negative and smaller hole effective mass value than its electron counterpart in direction $\Gamma-$$\mathrm{Z}$. Hence, we conclude that transitions labeled $E_{0,\bot}$ and $E_{1,\bot}$ should be modelled as $M_0$-type CPs, and transition labelled $E_{0,\parallel}$ will be modelled as $M_1$-type CP.


\begin{figure}[htbp]
\centering
\includegraphics[width=\linewidth]{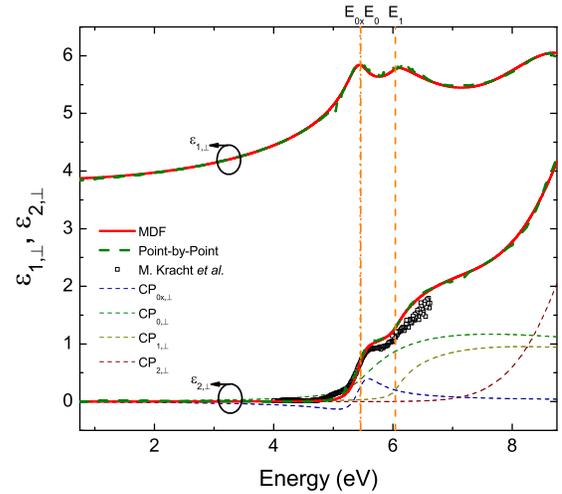}
\caption{Real and imaginary parts of the ordinary dielectric function, $\varepsilon_{\bot}$, for $\alpha$-Ga$_{2}$O$_3$ obtained by point-by-point analysis (PBP; dashed olive lines) and by MDF analysis (solid red lines). Note that PBP and MDF results are virtually indistinguishable. The individual CP contributions are shown for the imaginary part. Strong anharmonic exciton broadening is observed. Symbols indicate data from previous work by Kracht~\textit{et al.}, Ref.~\onlinecite{KrachtPRApplaGO2018}, measured on different samples. Vertical lines indicate MDF energy parameters given in Table~\ref{Tab:parms}.} 
\label{fig:epsordinary}
\end{figure}

\begin{figure}[htbp]
\centering
\includegraphics[width=\linewidth]{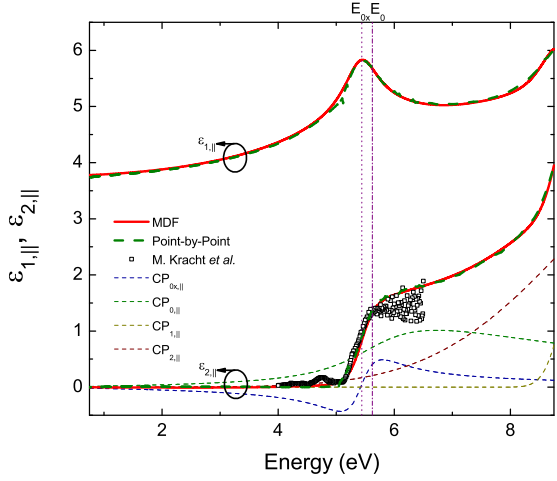}
\caption{Same as Fig.~\ref{fig:epsordinary} for $\varepsilon_{\parallel}$.} 
\label{fig:epsextraordinary}
\end{figure}

\begin{table}
\centering
\caption{Best-match CP MDF parameters and corresponding equation number from this paper for $\varepsilon_{\bot}$ and $\varepsilon_{\parallel}$. $\varepsilon_{\mathrm{off},\bot}$=1.647(5) and $\varepsilon_{\mathrm{off},\parallel}$=1.53(1). The last digit is determined with 90\% confidence, which is indicated with parentheses for each parameter.}
\begin{tabular}{{l}{c}{c}{c}{c}{c}{c}{c}}
\hline \hline
\multicolumn{6}{c}{$\varepsilon_{\bot}=\varepsilon_{\mathrm{off},\bot}+\varepsilon_{(M_0)}(E_0)+\varepsilon_{(ex)}(E_{0x})+\varepsilon_{(M_0)}(E_1)+\varepsilon_{(G)}(E_2)$}\\
\hline
&{CP}&Eq.~No.&{$A$}&{$E$ (eV)}&{$\Gamma$ (eV)}&{$b$ (eV)}\\
\hline
&CP$^{(0x)}$&4&0.82(8)&5.45(6)&0.46(1)&0.21(2)\\
&CP$^{(0)}$&1,2&50(4)&5.46(6)&0.28(3)&-\\
&CP$^{(1)}$&1,2&44(3)&6.041(5)&0.09(1)&-\\
&CP$^{(2)}$&5-7&175.63(1)&10.2(8)&2.54(1)&-\\
\hline
\multicolumn{6}{c}{$\varepsilon_{\parallel}=\varepsilon_{\mathrm{off},\parallel}+\varepsilon_{(M_1)}(E_0)+\varepsilon_{(ex)}(E_{0x})+\varepsilon_{(G)}(E_1)+\varepsilon_{(G)}(E_2)$}\\
\hline
&CP$^{(0x)}$&4&0.48(7)&5.44(2)&0.728(7)&0.67(1)\\
&CP$^{(0)}$&2,3&0.68(1)&5.62(2)&0.86(2)&-\\
&CP$^{(1)}$&5-7&2.64(6)&9.2(8)&0.78(2)&-\\
&CP$^{(2)}$&5-7&4.09(5)&11.2(3)&5.41(6)&-\\
\hline\hline
\label{Tab:parms} 
\end{tabular}
\end{table}

\begin{figure}[htbp]
\centering
\includegraphics[width=\linewidth]{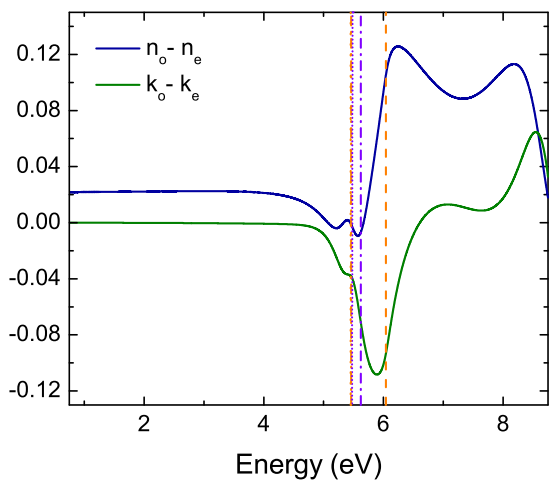}
\caption{Difference between ordinary and extraordinary refractive indices (blue line) and extinction coefficients (green line) from best-match MDF calculations. Vertical lines indicate the band-to-band transitions using the same line styles as in Figs.~\ref{fig:epsordinary} and~\ref{fig:epsextraordinary}. Note that $\alpha$-Ga$_2$O$_3$ is optically uniaxial negative in the below-bandgap spectral region.} 
\label{fig:MDFdifference}
\end{figure}

Figures~\ref{fig:epsordinary} and~\ref{fig:epsextraordinary} depict the best-match model calculated PBP (dashed lines) and MDF (solid lines) derived spectra for $\varepsilon_{\bot}$ and $\varepsilon_{\parallel}$, respectively. Both figures reveal the anisotropic ultra-wide bandgaps of $\alpha$-Ga$_2$O$_3$. Two distinct features can be identified in $\varepsilon_{\bot}$ in the near-bandgap region, where only one is seen in $\varepsilon_{\parallel}$. Also overlaid are data reported previously by Kracht~\textit{et al.} measured on different samples, which cover only parts of the second CP structure in $\varepsilon_{\bot}$.\cite{KrachtPRApplaGO2018}  In both $\varepsilon_{\bot}$ and $\varepsilon_{\parallel}$, features are broadened, similar to features within dielectric functions of other single crystalline ultra-wide bandgap metal oxides such as $\beta$-Ga$_2$O$_3$,\cite{SturmPRB2016,Mock_2017Ga2O3} ThO$_2$, or UO$_2$,\cite{MockAPLUO2THO22019} for example. We note that individual negative imaginary parts are due to the effect of anharmonic coupling. Separating a dielectric response into contributions from individual CP line shapes can result in limited spectral regions with some individual model functions revealing negative imaginary parts. This observation is a consequence of coupling between individual processes, such as between band-to-band transitions and exciton formation. Such observations are often made in line shape model approaches for the infrared optical properties in materials with coupled phonon modes.\cite{Gervais_1974} Energy conservation is valid for the sum of all present physical processes only. Table~\ref{Tab:parms} list all best-match model parameters, which together with equations given above suffice to near-perfect match the experimental data. By extrapolation of MDF values for $\hbar\omega\rightarrow 0$~eV we obtain $\varepsilon_{\infty,\bot}$ = 3.86 and $\varepsilon_{\infty,\parallel}$ = 3.76, which agree well with recent results obtained from infrared measurements ($\varepsilon_{\infty,\perp}$=3.75 and $\varepsilon_{\infty,\parallel}$=3.64).\cite{FenebergAPLaGOIR2019} We find the lowest band-to-band transitions for $\alpha$-Ga$_2$O$_3$ at $E_{0,\bot}$=5.46(6)~eV and $E_{0,\parallel}$=5.62(2)~eV, with $E_{0,\parallel} > E_{0,\bot}$. We find the binding energy parameter for the exciton contribution for the $M_0$-type transition, $E_{0,\bot}-E_{0x,\bot}$= 7~meV, which is much less than $E_{0,\parallel}-E_{0x,\parallel}$= 178~meV. Such small binding energy can be explained with the traditional exciton formation in a three dimensional effective hydrogen model. In this model, the ground state excitation equals the exciton binding energy ($R^{\star}$ = 13.6~eV$\mu/\varepsilon_{\mathrm{DC}}^2$), $\mu$ is the reduced mass of the combined density of states at the $\Gamma$-point, and $\varepsilon_{\mathrm{DC}}$ is the static dielectric constant. Using an estimate for $\varepsilon_{\mathrm{DC},\bot}$=10,\cite{SeguraPRMataGOEg2017} and reduced mass parameters listed in the caption of Table~\ref{tab:mass} we obtain $R^{\star}$~=~25~meV, in good agreement with our MDF result (7~meV). For the $M_1$-type CP, the exciton is hyperbolic and two-dimensional, and its ground state energy is four times smaller than that of a $M_0$-type exciton.\cite{Masaki2DJPSJ1966,Yu99} For large energy, short radius excitons phonon-exciton interaction is considered small.\cite{BechstedtExcitonBook} Therefore, $R^{\star}$ may be estimated using screening by $\varepsilon_{\infty}$ instead of $\varepsilon_{\mathrm{DC}}$. With our values for $\varepsilon_{\infty,\parallel}$ and mass parameters in Table~\ref{tab:mass}, we obtain $R^{\star}$~=~540~meV. This estimate is still in good qualitative agreement with our MDF result (178~meV). Our finding of such different excitons at the band edge of $\alpha$-Ga$_2$O$_3$ is not surprising given the highly anisotropic nature of both the real-space dielectric response as well as the reciprocal space band structure in the vicinity of the $\Gamma$-point. We note that Bechstedt and Furthm\"{u}ller recently estimated an exciton energy of 184~meV ignoring valence band contributions and anisotropy.\cite{doi:10.1063/1.5084324} We also note that the exciton broadening parameters are much larger than their binding energies. The influence of the correlated electron-hole pairs to the optical properties of semiconductors in the vicinity of the absorption threshold for the case of large broadening was discussed by Tanguy~\textit{et al.}\cite{TanguyPRB1999} It was shown analytically that a strong modification of the dielectric function in the near-band-to-band transition region is still present in such overdamped situations, even when room temperature thermal energy is larger than the exciton binding energy as observed here for the ordinary exciton. We further note that a similar CP analysis for monoclinic $\beta$-Ga$_2$O$_3$ resulted in exciton energy parameters of 120~meV, 230~meV, and 178~meV for the three fundamental band-to-band transitions polarized nearly along axes $c$, $a$, and $b$, respectively.\cite{Mock_2017Ga2O3} A similar observation was made by Sturm~\textit{et al.} who assumed equal energies for all transitions of 270~meV.\cite{SturmPRB2016} Additional research on the excitonic contributions to critical points, especially in complex low-symmetry materials such as transition metal oxides, will improve our understanding of their optical properties.
A higher energy transition is identified at $E_{2,\bot}$=10.2(8), which is obtained by extrapolation, i.e., through modeling of a Gaussian tail into the measured spectral range. This energy agrees well with a transition at 11.1~eV observed in synchrotron experiments.\cite{FenebergPRMataGOoDF2018} Two higher energy transitions are suggested in $\varepsilon_{\parallel}$ at 9.2(8)~eV and 11.2(3)~eV. Finally, Fig.~\ref{fig:MDFdifference} depicts differences between the ordinary and extraordinary indices of refraction and extinction coefficients, where it is seen that $\alpha$-Ga$_2$O$_3$ is uniaxial negative below the bandgap with very small and nearly wavelength independent birefringence ($\approx$0.02), and with small dichroism across the onset of absorption because both major directions absorb at about the same photon energy.    

In summary, we  have performed a combined ellipsometry and density functional theory study to determine the properties of the fundamental band-to-band transitions in $\alpha$-Ga$_2$O$_3$. We have identified and accurately modelled the contributions of three transitions at the onset of absorption, which (a) belong to two distinct three-dimensional van Hove singularities at maxima for transitions polarized perpendicular to the $c$ axis, and (b) to one $M_1$-type saddle point singularity for the transition polarized parallel to $c$ axis. Accordingly, we observe excitonic contributions with 7~meV binding energy for the lowest $M_0$-type transition, and with 178~meV binding energy for the $M_1$-type transition.

This work was supported in part by the National Science Foundation under award DMR 1808715, by Air Force Office of Scientific Research under award FA9550-18-1-0360, by the Nebraska Materials Research Science and Engineering Center under award DMR 1420645, by the Swedish Knut and Alice Wallenbergs Foundation supported grant 'Wide-bandgap semi-conductors for next generation quantum components', and by the American Chemical Society/Petrol Research Fund. M.~S. acknowledges the University of Nebraska Foundation and the J.~A.~Woollam~Foundation for financial support. DFT calculations were in part performed at the Holland Computing Center of the University of Nebraska, which receives support from the Nebraska Research Initiative.

The data that support the findings of this study are available from the corresponding author upon reasonable request.

\section{References}


%

\end{document}